\documentclass[aps,pra,twocolumn,showpacs,superscriptaddress,floatfix]{revtex4}
\usepackage{graphicx}
\usepackage{bm}
\begin{document}
\title{Ultracold RbSr molecules can be formed by magnetoassociation}
\author{Piotr S. \.Zuchowski}
\email{E-mail: Piotr.Zuchowski@durham.ac.uk}%
\affiliation{Department of Chemistry, Durham University, South
Road, Durham, DH1~3LE, United Kingdom}
\author{J. Aldegunde}
\email{E-mail: jalde@usal.es}%
\affiliation{Departamento de Qu{\'\i}mica F{\'\i}sica, Facultad de
Ciencias Qu{\'\i}micas, Universidad de Salamanca, 37008, Salamanca,
Spain}
\author{Jeremy M. Hutson}
\email{E-mail: J.M.Hutson@durham.ac.uk}%
\affiliation{Department of Chemistry, Durham University, South Road,
Durham, DH1~3LE, United Kingdom}

\date{\today}

\begin{abstract}
We investigate the interactions between ultracold alkali metal
atoms and closed-shell atoms using electronic structure
calculations on the prototype system Rb+Sr. There are molecular
bound states that can be tuned across atomic thresholds with
magnetic field, and there are previously neglected terms in the
collision Hamiltonian that can produce zero-energy Feshbach
resonances with significant widths. The largest effect comes
from the interaction-induced variation of the Rb hyperfine
coupling. The resonances may be used to form paramagnetic polar
molecules if the magnetic field can be controlled precisely
enough.
\end{abstract}

\pacs{34.50.Cx, 37.10.Pq, 67.85.-d}

\maketitle

There have been enormous advances in ultracold molecule
formation in the last few years. Alkali metal dimers have been
formed both by photoassociation and by magnetic tuning across
zero-energy Feshbach resonances (magnetoassociation). In both
approaches the molecules are initially formed in very
high-lying vibrational states, but methods are now emerging to
transfer the molecules to low-lying vibrational states
\cite{Sage:2005, Ni:KRb:2008, Lang:ground:2008, Danzl:v73:2008,
Viteau:2008, Deiglmayr:2008, Mark:2009, Haimberger:2009}. For
KRb and Cs$_2$, molecules in the rovibronic ground state have
been formed coherently from ultracold atoms by
magnetoassociation followed by stimulated Raman adiabatic
passage (STIRAP) \cite{Ni:KRb:2008, Danzl:ground:2010}. The
availability of samples of polar ultracold molecules opens up
new possibilities for exploring polar quantum gases, for
developing quantum simulators, and for studies of controlled
ultracold chemistry \cite{Carr:NJPintro:2009}.

Alkali metal dimers have singlet ground states, which cannot be
tuned magnetically except through the very small magnetic
moments of the constituent nuclei. Although the molecules can
be formed in excited triplet states, the triplet dimers are
likely to be subject to trap loss due to fast inelastic and
reactive collisions. There is therefore great interest in
forming ultracold molecules that have ground states with
unpaired electron spin. Molecules that have {\em both} electric
and magnetic dipoles are particularly interesting, because they
may be used to create topologically ordered states and may have
novel applications in quantum information storage
\cite{Micheli:2006}.

A very promising class of molecules are those formed from an
alkali metal atom, with a $^2$S ground state, and a
closed-shell species such as an alkaline earth atom or a Yb
atom, with a $^1$S ground state. However, the collision
Hamiltonian that is usually used for ultracold atom collisions
does not couple the atomic and molecular states in such
systems, and it is commonly believed that they will not exhibit
magnetically tunable Feshbach resonances that can be used for
ultracold molecule formation. The purpose of the present paper
is to show that this is not in fact true: there are terms in
the Hamiltonian that have previously been neglected, which can
give rise to magnetically tunable Feshbach resonances. If
precise enough control of magnetic fields can be achieved,
these resonances may be used for molecule formation in these
important systems.

We consider the prototype system RbSr. This is particularly
topical because Bose-Einstein condensation has recently been
achieved for $^{84}$Sr \cite{Stellmer:2009, Escobar:2009} and
$^{88}$Sr \cite{Mickelson:2010} and Fermi degeneracy for
$^{87}$Sr \cite{DeSalvo:2010}. Both $^{87}$Rb and $^{85}$Rb are
readily condensed. However, our general conclusions apply to
any system made up of an alkali metal atom and a closed-shell
species.

The collision Hamiltonian for a pair of atoms $a$ and $b$ is
\begin{equation}
\frac{\hbar^2}{2\mu}\left[-r^{-1}\frac{d^2}{dr^2}r +
\frac{\hat L^2}{r^2}\right] + \hat H_a + \hat H_b + \hat V(r),
\end{equation}
where $r$ is the internuclear distance, $\hat L^2$ is the
angular momentum operator for mechanical rotation of the atoms
about one another, $\hat H_a$ and $\hat H_b$ are the
Hamiltonians for the free atoms (in an applied field) and $\hat
V(r)$ is the interaction operator. For collision of a pair of
alkali metal atoms, $\hat H_a$ and $\hat H_b$ are
\begin{equation}
\hat H_\alpha = \zeta_\alpha \hat i_\alpha \cdot \hat s_\alpha + \left(
g_\alpha^e \mu_{\rm B} \hat s_{\alpha z} + g_\alpha^{\rm nuc} \mu_{\rm N} \hat
i_{\alpha z}\right) B,
 \label{eq:hmon}
\end{equation}
where $\zeta_\alpha$ is the hyperfine coupling constant for
atom $\alpha$, $\hat s_\alpha$ and $\hat i_\alpha$ are the
corresponding electron and nuclear spin operators, $g_\alpha^e$
and $g_\alpha^{\rm nuc}$ are the $g$-factors, and $B$ is the
magnetic field, whose direction defines the $z$-axis. The
interaction operator is usually represented
\begin{equation}
\hat V(r) = \sum_{S=|s_a-s_b|}^{s_a+s_b}
|S\rangle \langle S|\hat V|S\rangle \langle S| + \hat V^{\rm d}(r),
\label{eq:vint}
\end{equation}
where for a pair of alkali metal atoms $V_0(r) = \langle 0|\hat
V(r)|0\rangle$ and $V_1(r) = \langle 1|\hat V(r)|1\rangle$ are
the potential curves for the singlet and triplet states and
$\hat V^{\rm d}(r)$ is a spin-spin term that represents the
magnetic dipole interaction between the electron spins on the
two atoms (and may incorporate short-range terms due to
second-order spin-orbit interaction).

The molecular wavefunctions are conveniently expanded in an
uncoupled basis set $|s_am_{sa}\rangle |i_am_{ia}\rangle
|s_bm_{sb}\rangle$ $|i_bm_{ib}\rangle|LM_L\rangle$. The
Hamiltonian is diagonal in the total projection quantum number
$M_{\rm tot}=M_F+M_L$, where $M_F=m_{sa}+m_{ia}+m_{sb}+m_{ib}$.
The only term in the Hamiltonian that has matrix elements
off-diagonal in $L$ is the spin-spin term. However, for a pair
of alkali-metal atoms, the singlet potential is substantially
different from the triplet potential (typically a factor of 10
to 20 deeper), and the difference produces strong couplings
between states with the same $L$ and the same $M_F$. The
resulting molecular states typically have magnetic moments
different from the atomic states, and may cross them as a
function of magnetic field. The molecular states are coupled to
the atomic states by either $V_1(r)-V_0(r)$ (when the atomic
and molecular states have the same values of $L$ and $M_F$) or
$\hat V^{\rm d}(r)$ (when the $L$ or $M_F$ values are
different). This produces magnetically tunable Feshbach
resonances that may be used for molecule formation.

The situation is different when one of the atoms is in a $^1$S
state. When $s_b=0$, there is no spin-spin interaction. There
is also only one interaction potential, corresponding to
$S=s_a$ (a doublet, with $S=\frac{1}{2}$, for RbSr). The
quantum number $F=f_a=i_a\pm\frac{1}{2}$ is well-defined only
at zero field, but even at finite field the molecular states
have almost exactly the same mixture of $F$ values as the
atomic states. For the stable isotopes $^{84}$Sr, $^{86}$Sr and
$^{88}$Sr, which all have zero nuclear spin, the molecular
states have almost exactly the same magnetic moment as their
parent atomic states, so are closely parallel to them as a
function of magnetic field. Even for $^{87}$Sr, with $i_b=9/2$,
the nuclear Zeeman effect is a small perturbation.

If the interaction operator was really represented by Eq.\
(\ref{eq:vint}), with only $S=\frac{1}{2}$ and no $\hat V^{\rm
d}(r)$, there would be no coupling between the atomic and
molecular states. Although there would still be crossings
between atomic states and molecular bound states as a function
of magnetic field, there would be no coupling between them; the
resulting Feshbach resonances would have zero width and it
would be impossible to tune across them adiabatically, as
required for molecule formation. However, Eq.\ (\ref{eq:vint})
is in reality an approximation, and there are several
additional ways in which the colliding species interact with
one another. For RbSr, these additional terms have a profound
effect.

The most important additional interaction term comes from the
fact that the Rb hyperfine coupling constant $\zeta$ is
modified when another atom is nearby. We may write $\zeta(r) =
\zeta_{\rm Rb} + \Delta\zeta(r)$. The term $\Delta\zeta(r) \hat
i_a\cdot\hat s_a$ is most appropriately considered to be part
of the interaction operator $\hat V(r)$. In addition, there are
short-range contributions to $\hat V(r)$ from smaller terms
such as (i) the interaction $eQq(r)$ between the nuclear
quadrupole moment of Rb and the field gradient at the Rb
nucleus caused by the distortion of the electron density by Sr;
(ii) the dipolar interactions between the electron spin and the
Rb nuclear spin; (iii) the spin-rotation interactions between
$L$ and the Rb electron and nuclear spins. However, the
modification of $\zeta_{\rm Rb}$ is by far the largest effect,
as discussed below.

\begin{figure}
\includegraphics[scale=0.9]{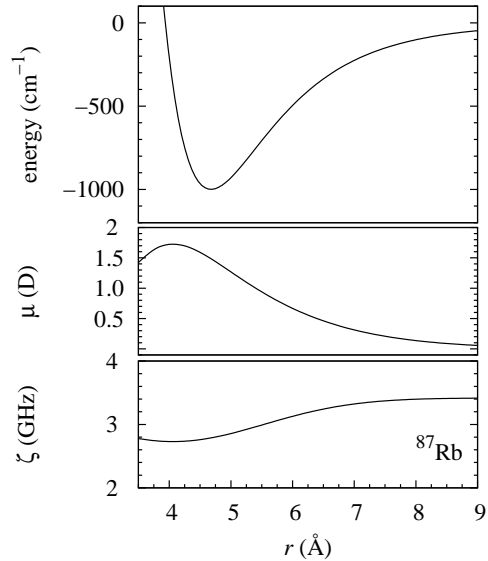}
\caption{Interaction potential $V(r)$ (top panel) with dipole
moment $\mu(r)$ (middle panel) and hyperfine coupling constant
$\zeta(r)$ (bottom panel). The binding energy and equilibrium
distance are calculated to be 1000 cm$^{-1}$ and 4.67 \AA,
respectively.} \label{pots}
\end{figure}

We have carried out high-level electronic structure
calculations of the interaction potential and dipole moment of
RbSr and the modification of $\zeta_{\rm Rb}$ by Sr. The
results are summarized in Fig.\ \ref{pots}. The short-range
potential (from 3 to 10 \AA) was calculated using the
spin-restricted coupled-cluster method with single, double and
approximate triple excitations [CCSD(T)], with the relativistic
small-core ECP28MDF effective core potentials and uncontracted
basis sets of Lim {\em et al.}\ \cite{Lim:2005, Lim:2006}
augmented by $3s3p2d$ functions at the bond midpoint. For
$R\!>\!10$~\AA, we extrapolate using the long-range form
$V(r)=-C_6r^{-6}-C_8r^{-8}$, with semiempirical coefficients
\cite{Mitroy:2003} $C_6=3.762\times10^3\ E_{\rm h} a_0^6$ and
$C_8=4.62\times10^5 E_{\rm h} a_0^8$. The short- and long-range
parts of the potential were smoothly connected with the
switching function used by Janssen {\em et al.}
\cite{Janssen:2009}. The quantities $\zeta(r)$ and $eQq(r)$
were calculated with the relativistic density-functional theory
(DFT) approach \cite{vanLenthe:1994} implemented in the ADF
program \cite{ADF1}, using the PBE0 functional
\cite{Adamo:1999}. The asymptotic value of $\zeta_{\rm Rb}$ was
underestimated by 6\% in the DFT calculations, so we scaled
$\zeta(r)$ to reproduce the experimental atomic value. The
quantity $\Delta\zeta(r)$ was fitted to the Gaussian form
$\zeta_0 e^{-a(r-r_c)^2}$, giving parameters $a=0.23$
\AA$^{-2}$ and $r_c=4.06$~\AA, with $\zeta_0=-687$ MHz for
$^{87}$Rb and $-229$ MHz for $^{85}$Rb. This corresponds to a
20\% maximum reduction in $\zeta$. We also evaluated the dipole
moment $\mu(r)$ using a variety of approaches. The most
reliable is a finite-field CCSD(T) calculation with the
approach described above, which gives $\mu=1.36$~D at the
equilibrium distance $r_e=4.67$ \AA. However, the precise value
is sensitive to the level of correlation treatment.

We have also investigated the smaller additional couplings
described above, in order to verify that they are much less
important than $\Delta\zeta(r)$. All these can couple states
with $\Delta M_F\ne0$ when $L>0$. The nuclear quadrupole
coupling and the dipolar coupling between $s_a$ and $i_a$ can
also couple channels with $\Delta L\ne0$ and can thus mediate
Feshbach resonances in s-wave scattering for bound states with
$L\ne0$. Our DFT calculations give values for the nuclear
quadrupole coupling coupling constant $eQq(r_e)=8$~MHz and
3.7~MHz for $^{85}$RbSr and $^{87}$RbSr respectively, reducing
rapidly to zero as the RbSr distance increases beyond $r_e$.
The coupling constant for the dipolar interaction between $s_a$
and $i_a$ is of the order of 1 MHz near $r_e$. The electronic
spin-rotation coupling constant can be estimated in terms of
the rotational constant $b$ and the anisotropy $g_{\rm aniso}$
of the electronic $g$-factor as $2b g_{\rm aniso}$
\cite{Curl:1965}, which is approximately 20~MHz near $r_e$. The
nuclear spin-rotation interaction will be about a factor of
$10^3$ smaller because of the ratio of the nuclear and Bohr
magnetons. A 20~MHz coupling is potentially significant, but
neither the electronic nor the nuclear spin-rotation
interaction has matrix elements that affect s-wave scattering.
All these smaller coupling terms, and couplings involving the
nuclear spin of $^{87}$Sr, are neglected in the scattering
calculations described below.

\begin{figure}
\includegraphics[scale=0.65,angle=270]{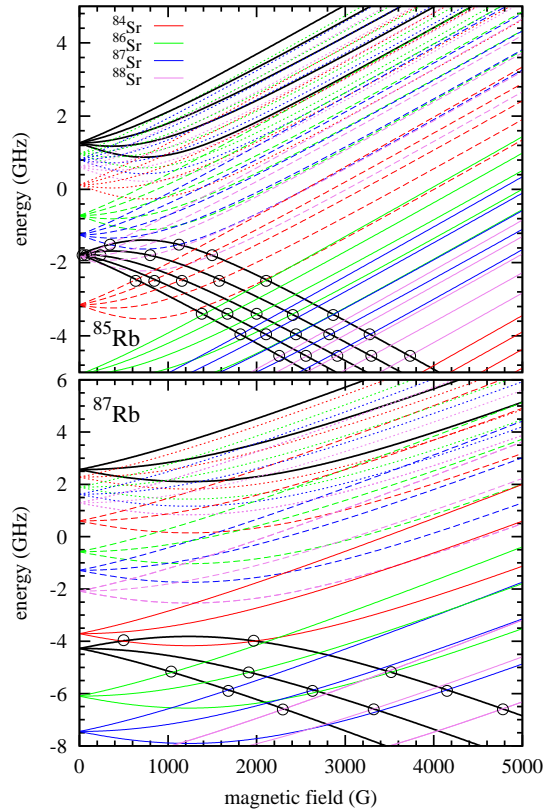}
\caption{(Color online) Molecular levels (colored) that cross
atomic thresholds (black) for $|M_F|\leq i_a-\frac{1}{2}$ states
of $^{85}$RbSr (upper panel) and $^{87}$RbSr (lower panel) as a
function of magnetic field for different Sr isotopes. The
dotted, dashed and solid colored lines correspond to
$v=-2,-3,-4$ vibrational states, respectively. Positions of
resonances are marked with circles.
} \label{thresh}
\end{figure}

\begin{table}
\caption{Calculated properties of RbSr Feshbach resonances at
$F=i_a-\frac{1}{2}$ thresholds arising from bound states
supported by the $F=i_a+\frac{1}{2}$ thresholds.
\label{tab:res}}
\begin{ruledtabular}
\begin{tabular}{lcccc}
 system &    $B_{\rm res}$ (G)  &  $a_{\rm bg}$ (\AA) & $\Delta$ (mG) & $M_F$  \\ \hline
$^{85}$Rb$^{84}$Sr &   632            &   19.5      &  -0.0560      &  +2     \\
         &   840            &   19.5      &  -0.292       &  +1     \\
         &   1145           &   19.5      &  -0.636       &   0     \\
         &   1562           &   19.5      &  -0.987       &  -1     \\
         &   2076           &   19.4      &  -0.636       &  -2     \\    \hline
$^{85}$Rb$^{86}$Sr &   1383           &   815.7     &  -1.83        &  +2     \\
         &   1649           &   813.9     &  -4.34        &  +1     \\
         &   1977           &   814.1     &  -7.14        &   0     \\
         &   2371           &   814.1     &  -8.85        &  -1     \\
         &   2827           &   814.1     &  -7.50        &  -2     \\    \hline
$^{85}$Rb$^{87}$Sr &     336          &   95.1      &   0.0545      &  -2      \\
         &     1108         &   95.1      &  -0.586       &  -2     \\
         &     1797         &   95.1      &  -0.163       &  +2     \\
         &     2079         &   95.1      &  -0.347       &  +1     \\
         &     2413         &   95.1      &  -0.532       &   0     \\
         &     2800         &   95.1      &  -0.664       &  -1     \\
         &     3240         &   95.0      &  -0.499       &  -2     \\ \hline
$^{85}$Rb$^{88}$Sr &     37           &   56.2      &  -0.000277    &  +2      \\
         &     70           &   56.2      &  -0.00239     &  +1     \\
         &     235          &   56.2      &  -0.0397      &   0     \\
         &     792          &   56.2      &  -0.270       &  -1     \\
         &    1481          &   56.2      &  -0.311       &  -2     \\
         &    2237          &   56.2      &  -0.124       &  +2     \\
         &    2531          &   56.2      &  -0.276       &  +1     \\
         &    2869          &   56.2      &  -0.402       &   0     \\
         &    3253          &   56.2      &  -0.485       &  -1      \\
         &    3680          &   56.2      &  -0.350       &  -2     \\     \hline
$^{87}$Rb$^{84}$Sr &   477            &   1715.0    &   7.41        &  -1     \\
         &   1959           &   1700.3    &  -122         &  -1     \\         \hline
$^{87}$Rb$^{86}$Sr &   1036           &   55.0      &  -0.209       &  +1     \\
         &   1896           &   55.0      &  -1.08        &   0     \\
         &   3472           &   55.0      &  -2.29        &  -1     \\     \hline
$^{87}$Rb$^{87}$Sr &   1660           &   31.5      &  -0.636       &  +1     \\
         &   2608           &   31.5      &  -2.27        &   0     \\
         &   4096           &   31.5      &  -3.79        &     -1     \\     \hline
$^{87}$Rb$^{88}$Sr &   2281           &    1.6      &  -33.6        &     +1     \\
         &   3280           &    1.6      &  -101         &      0      \\
         &   4716           &    1.5      &  -153         &     -1      \\
\end{tabular}
\end{ruledtabular}
\end{table}

Since the $M_F$-changing terms in the collision Hamiltonian are
so small, crossings between thresholds and bound states with
different $M_F$ will not produce Feshbach resonances wide
enough to be measured with current experimental methods.
However, if the Rb atom is initially in a state that correlates
at zero field with $F=i_a-\frac{1}{2}$, then the threshold is
crossed by bound states with the same $M_F$ but correlating
with $F=i_a+\frac{1}{2}$. The operator $\Delta\zeta(r) \hat
i_a\cdot\hat s_a$ is {\em not} diagonal in the field-dressed
atomic eigenstates, so it can produce Feshbach resonances at
these crossings. The pattern of bound states and the crossings
that produce Feshbach resonances are shown in Fig.\
\ref{thresh} for $^{85}$Rb and $^{87}$Rb with all the stable
isotopes of Sr. Since the hyperfine splittings are a few GHz,
the bound states responsible for the crossings are bound by
energies of a few GHz. For RbSr these are states with
vibrational quantum numbers $v=-3,-4$ (relative to threshold)
for magnetic fields up to 0.5~T.

We next investigated the widths of the resonances produced in
this way. To locate them, we first carried out bound-state
calculations as a function of magnetic field using the BOUND
package \cite{Hutson:Cs2:2008} to determine precisely the
magnetic field at which the crossing occurs. We then used the
MOLSCAT program \cite{molscat:v14}, modified to handle
collisions of atoms in magnetic fields
\cite{Gonzalez-Martinez:2007}, to calculate the scattering
length $a(B)$ as a function of magnetic field near the crossing
at a near-zero collision energy ($10^{-8}$~K). This was then
fitted to the functional form $a(B)=a_{\rm bg}\left[1 + \Delta
/ (B - B_{\rm res})\right]$ to extract the resonance position
$B_{\rm res}$, the width $\Delta$, and the background
scattering length $a_{\rm bg}$.

The parameters of selected Feshbach resonances are given in
Table \ref{tab:res}.
It should be noted that our calculated interaction potential is
not accurate enough to predict the positions of the highest
bound states (or the scattering length) correctly for a
specific isotope. However, there are enough different isotopes
of Sr available that one of them is likely to display crossings
of each of the types shown in Fig.\ \ref{thresh}. A measurement
of the scattering length or the binding energy of one of the
near-dissociation states for any isotopic species will allow
reliable calculations of the resonance positions for all
isotopic combinations.

States with $M_F<0$ (except $M_F=-i_a-\frac{1}{2}$) have energy
maxima or minima at $B_{\rm turn} =-\zeta M_F/(g_e \mu_{\rm
B})$, where they are separated by $d = \zeta\left[
(i_a+\frac{1}{2})^2-M_F^2\right]^{1/2}$. For levels with
binding energies $|E_v|$ in the range $ d < E_v <
(i_a+\frac{1}{2})\zeta$, there are two crossings with the lower
threshold: one between 0 and $B_{\rm turn}$ and the second
between $B_{\rm turn}$ and 2$B_{\rm turn}$. Conversely, for
levels with binding energies $|E_v| > (i_a+\frac{1}{2})\zeta$,
there is only one crossing, at a field $B>2B_{\rm turn}$. For
levels with $M_F\geq 0$, there is only one crossing for each
vibrational state with binding energy larger than
$(i_a+1/2)\zeta$.

The resonance width is proportional to the square of a
bound-continuum matrix element $\langle v | \Delta\zeta(r) \hat
i_a\cdot\hat s_a| \hbox{continuum}\rangle$. The radial part of
this is proportional to the amplitude of the bound-state
wavefunction at short range, which varies as
$|E_v|^{(n+2)/4n}$, where $n=6$ is the power of the leading
term in the long-range potential. Since $\Delta\zeta(r)$ is
itself roughly proportional to $\zeta$, the widths of the
low-field resonances may generally be expected to increase with
$\zeta$ approximately as $\zeta^{8/3}$. Because of this, the
widths observed for $^{87}$Rb in Table \ref{tab:res} are
generally larger than those for $^{85}$Rb. However, there are
other factors involved such as the likelihood of obtaining
low-field resonances at all (which decreases with $\zeta$) and
the magnitude of the off-diagonal matrix element of $\hat
i_a\cdot\hat s_a$, which increases linearly with $B$ at low
field and then levels off above $B_{\rm turn}$ to a value
proportional to $\left[(i_a+\frac{1}{2})^2-M_F^2\right]^{1/2}$.
Because of this, the lowest-field resonances in Table
\ref{tab:res} are all very narrow. Resonance widths are also
enhanced in cases where the background scattering length is
large.

The widest and most promising resonances for molecule
production are of two types. The first are those that occur for
$M_F<0$ states that cross the lower threshold twice, as occurs
for $v=-3$ for $^{85}$Rb$^{87}$Sr and $v=-4$ for
$^{87}$Rb$^{84}$Sr in Fig.\ \ref{thresh}. The higher-field
resonance is always the wider of the pair because of the $\hat
i_a\cdot\hat s_a$ matrix element discussed above. The second
are those for which the background scattering length is large,
as for $^{85}$Rb$^{86}$Sr and $^{87}$Rb$^{84}$Sr on the current
potential. These two effects combine for the 1959 G resonance
for $^{87}$Rb$^{84}$Sr in Table \ref{tab:res} to give a width
as high as 122 mG.

In conclusion, we have investigated the interactions between Rb
and Sr atoms and have identified a new mechanism that can
produce magnetically tunable Feshbach resonances in collisions
of ultracold molecules. These Feshbach resonances could be used
to produce ultracold molecules that would have both electric
and magnetic dipole moments in their ground states. The
resonances arise from the modification of the Rb hyperfine
coupling by the presence of another atom. The mechanism is
general and may produce magnetically tunable Feshbach
resonances in any system in which an atom with electron spin
collides with a closed-shell atom.

This work is supported by EPSRC under collaborative projects
CoPoMol and QuDipMol of the ESF EUROCORES Programme EuroQUAM.

\bibliography{../../all}

\end{document}